\documentclass[reprint,amsmath,amssymb,showkeys,aps,prb]{revtex4-1}
\usepackage{natbib}
\usepackage{graphicx}
\usepackage{bm}
\usepackage{hyperref}

\begin{document}
\title{Microscopic Friction Emulators}
\author{Davide Mandelli$^{1}$ \& Erio Tosatti$^{1,2,3}$}
\affiliation{
\vspace{5pt}
$^1$ International School for Advanced Studies (SISSA), Via Bonomea 265, 34136 Trieste, Italy \\
$^2$ CNR-IOM Democritos National Simulation Center, Via Bonomea 265, 34136 Trieste, Italy \\
$^3$ International Centre for Theoretical Physics (ICTP), Strada Costiera 11, 34014 Trieste, Italy\\
            }
\date{\today}
\begin{abstract}
Cold ions sliding across periodic potential patterns formed by lasers elucidate the physics of dry friction 
between crystals. Experiments with no more than six ions suffice to explore a vast domain of frictional forces.
\end{abstract}
\maketitle

 \begin{figure*}[!t]
 \begin{center}
 \includegraphics[angle=0, width=1\textwidth]{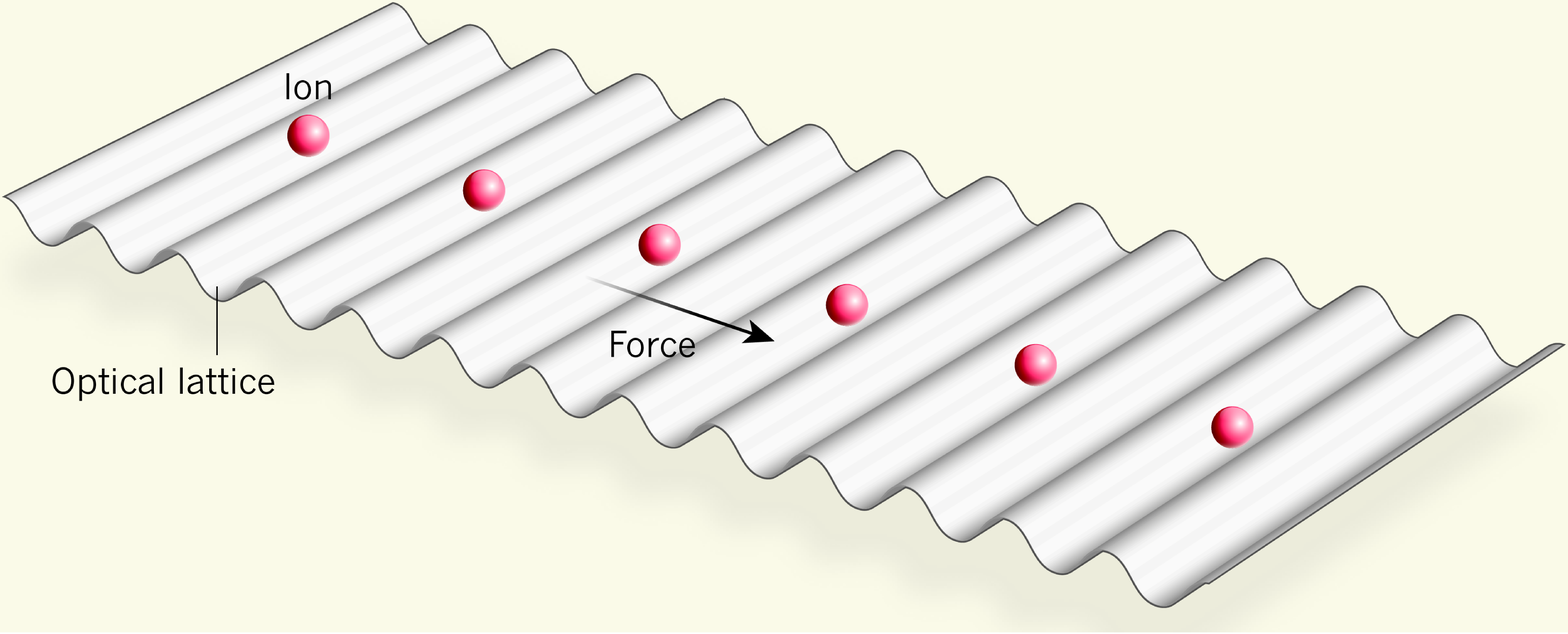}
 \caption{ \label{fig1} 
  {\bf Ion chains trapped in a lattice.} Bylinskii et al.~\cite{Bylinskii} and Gangloff et al.~\cite{Gangloff} 
  investigate the frictional forces arising when one or a few ions are forced to slide across optical 
  lattices, where the corrugated intensity potentials  is created by counter-propagating lasers. 
  These simple experiments have provided benchmark tests of  long-standing frictional models for 
  sliding crystal surfaces.}
 \end{center}
 \end{figure*}

The study of friction is a centuries-old, but still vibrant, subject. The simplest, archetypal 
dry sliding friction problem is that between crystal surfaces, more precisely the force resisting 
the relative lateral motion of two crystals in contact. In this case, the periodic spacing of the 
atoms in the crystal eliminates the complexity introduced by the ill-defined nature of  ordinary, 
non-crystalline surfaces. Studies of nanometre-scale systems through non-equilibrium theory, computer 
simulations, and most recently the use of artificial friction emulators have revived this topic.~\cite{Vanossi}
Writing in Science and in Nature Physics, respectively, Bylinskii et al.~\cite{Bylinskii} and 
Gangloff et al.~\cite{Gangloff} follow up on earlier theoretical suggestions,~\cite{Benassi,GarciaMata,Pruttivarasin} 
and present friction emulators formed from trapped ions that slide over optical lattices; the latter are 
periodic intensity potentials created by the interference of laser beams.

Although dry sliding is a well-defined problem, sufficient inroads have not yet been made into it -- microscopically well 
defined experimental systems are not abundant. Theory of frictional dynamics is often limited to computer 
simulations -- vivid but often hasty and incomplete representations of the real, complex and slow sliding 
of true solid interfaces. Much of our understanding relies on apparently trivial one-dimensional periodic 
potential models, such as the Prandt-Tomlinson (PT) single-slider~\cite{Muser} and the Frenkel-Kontorova (FK) 
sliding-chain descriptions.~\cite{Krylov} The PT model illustrates the switch from stick-slip  (the typical 
stop-and-go advance of, say, a chalk stick on a school blackboard) to smooth sliding for decreasing 
surface corrugation, as well as the passage from large friction at high sliding speeds to vanishingly 
small friction (thermolubricity) at low speeds.~\cite{Muser,Krylov} In the special case of mismatched lattices, 
the FK model describes the transition between frictionless (superlubric) motion and pinned frictional sliding 
(the onset of static friction).

What these models have achieved is to describe how properties  such as corrugation, temperature, velocity, and 
lattice-matching of crystals in contact may influence and determine friction, and thus in principle permit its 
control, desirable as that is in many practical instances.
Despite much theoretical background,~\cite{Muser,Krylov} neither of these models has been really tested 
experimentally. That is what emulators can do.

The emulators of Ref.~\onlinecite{Bylinskii,Gangloff} are based on short chains of 
trapped ionized atoms that, under the influence of an electric field, are forced to slide across a 
laser-generated optical lattice. These techniques may seem arcane, yet they are accurate and powerful, 
because parameters such as the temperature, atom velocity and spacing, chain length and the amplitude of 
the lattice’s potential, can be flexibly adjusted across a vast range of values. 

In this vein, Gangloff and colleagues present an experiment involving mostly a single ion that slides 
across an optical lattice, a set-up that emulates the PT model to near perfection. Through it they 
demonstrate that,  even at their microKelvin temperature, there is a speed below which the friction 
between the sliding ion and the lattice vanishes (thermolubricity),  as demanded by thermodynamics. 
Above that speed, the friction reverts to stick-slip and rises above two orders of magnitude with 
increasing speed.  No experiment involving real crystals can span this kind of range.  Although much 
of the authors’ findings had been known from numerical simulations,~\cite{Muser,Krylov,Braun} their 
emulator is superior -- for example it reproduces much better the theoretically expected dependence 
of friction on velocity.~\cite{Sang}

Emulating the FK model after the PT model would  require an infinitely long ion chain in place of a single ion.  
Bylinskii et al. used  instead a short chain of two to six ions sliding across an optical lattice -- so is this 
just a baby step towards that goal? Not quite. By adjusting the distances between the ions, the authors tune 
the amount of mismatch between the chain and the corrugation of the lattice’s periodic potential, and produce 
a dramatic effect on the friction that they measure.

Starting from large friction for perfect chain-lattice matching, Bylinskii and colleagues observe a rapid 
decrease of the friction as they increase a parameter that controls the amount of chain-lattice mismatch. 
This trend reflects the evolution from strong pinning friction, a regime where it takes a large force to 
dislodge and nudge the chain forward, towards better lubricity, the sliding behaviour of  two mismatched 
lattices, whose relative position is energetically more indifferent. Thus for increasing mismatch the system 
tends towards lubricity; even if not necessarily superlubricity, a regime only reached  when the lattice 
potential intensity falls below the value that demarcates the  “Aubry” transition between frictionless and 
pinned frictional sliding.~\cite{Aubry} It will be interesting in the future to pursue the vestiges of this 
important transition,  known to survive even for very short chains,~\cite{Benassi,Sharma} as is the case in 
this kind of emulators. It is amazing to remark how much already at this stage the study of just a few ions 
can teach us about the physics of infinite systems.

As always, experiments teach us more than we anticipate. Clear-cut techniques, like the cold-ion emulators 
reported by these two research papers, provide useful insights into the complexity underlying even the simplest 
act of friction involving a handful of ions. In this sense, paraphrasing the ''more is different'' dictum by 
physicist Philip Warren Anderson,~\cite{Anderson} one could here say that sometimes ''less in not that 
different'' after all.

This work was mainly supported under the ERC Advanced Grant No.\ 320796-MODPHYSFRICT.



\begin{thebibliography}{99}

\bibitem{Vanossi} A. Vanossi, N. Manini, M. Urbakh, S. Zapperi, and E. Tosatti, Rev. Mod. Phys. {\bf 85}, 529 (2013).

\bibitem{Bylinskii} A. Bylinskii, D. Gangloff, V. Vuletic, Science {\bf 348}, 1115-1118 (2015).

\bibitem{Gangloff} D. Gangloff, A. Bylinskii, I. Counts, W. Jhe, V. Vuletic, Nat. Phys. {\bf 11}, 915-919 (2015).

\bibitem{Benassi} A. Benassi, A. Vanossi, E. Tosatti, Nat. Comm. {\bf 2}, 236 (2011).

\bibitem{GarciaMata} I. Garcia-Mata, O. V. Zhirov, D. L. Shepelyansky, Eur. Phys. J. D {\bf 41}, 325 (2007).

\bibitem{Pruttivarasin} T. Pruttivarasin, M. Ramm, I. Talukdar, A. Kreuter, H. H\"affner, New J. Phys. {\bf 13}, 075012 (2011) .

\bibitem{Muser} M. H. M\"user, Phys. Rev. B {\bf 84}, 125419 (2011).

\bibitem{Krylov} S. Y. Krylov, J. W. M.  Frenken, Phys. Status Solidi B {\bf 251}, 711 (2014).

\bibitem{Braun} O. M. Braun, Y. Kivshar, ''The Frenkel-Kontorova Model: Concepts, Methods and applications'' (Springer, Berlin, 1998).

\bibitem{Sang} Y.  Sang, M. Dub\'e, M. Grant, Phys. Rev. Lett. {\bf 87}, 174301 (2001).

\bibitem{Aubry} S. Aubry, P. Y. Le Daeron, Physica D {\bf 8}, 381 (1983).

\bibitem{Sharma} S. R. Sharma, B. Bergersen, B. Joos, Phys. Rev. B {\bf 29}, 6335 (1984).

\bibitem{Anderson} P. W. Anderson, Science {\bf 177}, 393 (1972).

\end{thebibliography}
\end{document}